\documentclass[aps,prl,twocolumn, superscriptaddress]{revtex4}
\usepackage{graphicx} 
\begin{document}

\title{Sub-100 attoseconds Optics-to-microwave synchronization}

\author{W. Zhang}
\affiliation{LNE-SYRTE, Observatoire de Paris, CNRS, UPMC, 75014 Paris, France}
\author{Z. Xu}
\affiliation{FEMTO-ST, Time and Frequency dept., CNRS, ENSMM, Besan\c{c}on, France}
\author{M. Lours}
\affiliation{LNE-SYRTE, Observatoire de Paris, CNRS, UPMC, 75014 Paris, France}
\author{R. Boudot}
\affiliation{FEMTO-ST, Time and Frequency dept., CNRS, ENSMM, Besan\c{c}on, France}
\author{Y. Kersale}
\affiliation{FEMTO-ST, Time and Frequency dept., CNRS, ENSMM, Besan\c{c}on, France}
\author{G. Santarelli}
\affiliation{LNE-SYRTE, Observatoire de Paris, CNRS, UPMC, 75014 Paris, France}
\author{Y. Le Coq}
\email{yann.lecoq@obspm.fr}
\affiliation{LNE-SYRTE, Observatoire de Paris, CNRS, UPMC, 75014 Paris, France}

\date{\today}

\begin{abstract}
We use two fiber-based femtosecond frequency combs and a low-noise carrier suppression phase detection system to characterize the optical to microwave synchronization achievable with such frequency divider systems. By applying specific noise reduction strategies, a residual phase noise as low as -120\,dBc/Hz at 1\,Hz offset frequency from a 11.55\,GHz carrier is measured. The fractional frequency instability from a single optical-to-frequency divider is $1.1 \times  10^{-16}$ at 1\,s averaging down to below $2 \times 10^{-19}$ after only 1000\,s. The corresponding rms time deviation is lower than 100 attoseconds up to 1000\,s averaging duration.
\end{abstract}

\maketitle

%---------------------------------------------------------------------
% Introduction
%---------------------------------------------------------------------

High-performance systems allowing the generation of ultra-stable microwave signals from optical sources are of great interest in a variety of applications such as radar, telecommunication and deep space navigation systems, timing distribution and synchronization \cite{Kim_Nature08, Kim_OL_07}, ultra-high resolution Very Long-Baseline Interferometry, and development of local oscillators for accurate fountain atomic frequency standards \cite{Weyers_PRA_09, Millo_APL_09}. 

Femtosecond lasers have revolutionized the field of time and frequency metrology since they appeared early as innovative and high-performance tools to realize a phase coherent link between optical and microwave frequencies \cite{Holzwarth_PRL_00, Ramond_OPL_02}. The conversion from the optical to the microwave domain is based on the synchronization of the pulse repetition rate on the optical frequency of an ultra-stable continuous (CW) laser and the subsequent detection of the optical pulse train, typically using a fast photodiode. This process is, however, accompanied by excess phase noise which limits the residual timing stability of the microwave frequency synthesis \cite{McFerran_EL_05, Bartels_OL_05, Ivanov_UFFC_07}.

In this letter, we implement advanced noise reduction techniques on a fiber-based optical frequency comb (FOFC) to synthesize microwave signals (at 11.55\,GHz) with an additive phase noise of -120\,dBc/Hz at 1\,Hz and $\sim -135$\,dBc/Hz at Fourier frequencies above 100\,Hz. The fractional frequency stability reads $\sim 1.1 \times 10^{-16}\tau^{-1}$ (where $\tau$ is the averaging time) up to 1000\,s. Our experiment demonstrates timing synchronization between the reference optical frequency and the synthesized microwave signal below 100\,as on this timescale. 

%---------------------------------------------------------------------
% Phase-lock and microwave generation scheme
%---------------------------------------------------------------------

The setup for phase-locking our FOFC on an ultra-stable laser, which allows the frequency synthesis of low noise microwave signals from the optical domain has been described previously in \cite{Millo_OL_09, Millo_APL_09}. Briefly, a 30\,mW output of a commercial FOFC (repetition rate $f_{\rm rep}\sim 250$\,MHz) is filtered near 1542\,nm in an Optical Add and Drop Module (OADM). The 0.8\,nm large first OADM's output is mixed with a CW reference of optical frequency $\nu_{\rm CW}$ leading to a beatnote $f_{\rm b}$. A built-in f-2f interferometer unit produces the carrier-envelop offset frequency $f_0$, which is mixed with $f_{\rm b}$ to produce a $f_0$ independent signal $\nu_{\rm CW}-N\times f_{\rm rep}$. After 64-fold digital division and comparison with a radio-frequency (RF) reference, this produces an error signal which is used to correct the repetition rate of the FOFC by acting on its pump diodes' current for fast Fourier frequencies \cite{Millo_OL_09, McFerran_APB_07} (above $\sim 2$\,kHz), and a cavity-length modifying piezoelectric actuator (PZT) for slow Fourier frequencies (below $\sim 2$\,kHz). To generate microwave signals, the second output of the OADM containing almost all of the optical spectrum is detected by a 20\,GHz-bandwidth InGaAs pigtailed photodiode. The output signal of the photodetector contains all the harmonics of $f_{rep}$ up to 20\,GHz. The available microwave power for the harmonics near 10\,GHz is approximately -30\,dBm per line. A bias-tee connected to the output of the photodiode allows monitoring on its DC port of the generated averaged electric power, while the high frequencies are directed to a microwave signal phase detection system. Two quasi-identical systems, one with $\sim 150$\,kHz total feed-back bandwidth, the other with only $\sim 20$\,kHz (due to different driving electronics) are phase-locked to the same ultra-stable optical reference. We assume the two FOFC systems to be statistically uncorrelated. As a consequence, the two microwave signals are phase compared to characterize the optical-to-microwave synchronization level of a single system. 

%---------------------------------------------------------------------
% Carrier supression phase detection scheme
%---------------------------------------------------------------------

\begin{figure}[ht]
\includegraphics[width=\columnwidth]{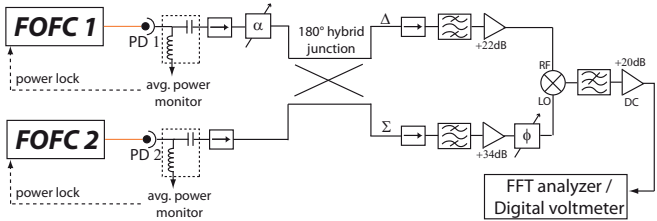}
\caption{Setup Schematic for a carrier suppression phase detection scheme between two FOFC generated microwave signals.}
\label{fig:SetUp_2FOFC_2PhD}
\end{figure}

To explore the ultimate performance of our optics to microwave synchronization system, we need to use an ultra-high sensitivity phase comparison scheme which we realize with a carrier suppression noise measurement system (CSNMS) \cite{Ivanov_IEEEUFFC_98, Rubiola_RSI_99} as shown on Fig. \ref{fig:SetUp_2FOFC_2PhD}. The electric signals resulting from pulse trains photodetections are directed to the two input ports of a 180-degrees hybrid junction through microwave isolators preventing unwanted feed-back effects on the photodiodes. A variable attenuator in front of one of these ports allows amplitude equalization of the signals. When phase matching is realized for a given carrier frequency, the difference output port of the hybrid junction exhibits carrier suppression and every remaining signal near the carrier frequency is resulting from phase or amplitude differential noise at the input ports. Note that phase matching can conveniently be adjusted by changing the phase of the RF reference in one of the FOFC's phase-lock loops. The sum and difference output ports of the hybrid junction are filtered near 11.55\,GHz (harmonic of interest throughout this paper) with low-insertion loss ultra-narrow cavity filters. The difference signal is afterward amplified and sent to the RF port of a microwave mixer. The sum signal passes through a variable phase shifter, is amplified and saturates the LO input port of the mixer. The output of the mixer is low-pass filtered, amplified by a low-noise DC amplifier and sent to a fast Fourier transform (FFT) analyzer or digital voltmeter acquisition platform. Depending on the relative phase between the dark and bright port on the mixer, the near-DC output of the mixer is proportional either to the phase or amplitude difference between the 11.55\,GHz harmonics at the input of the hybrid junction.

\begin{figure}[ht]
\includegraphics[width=\columnwidth]{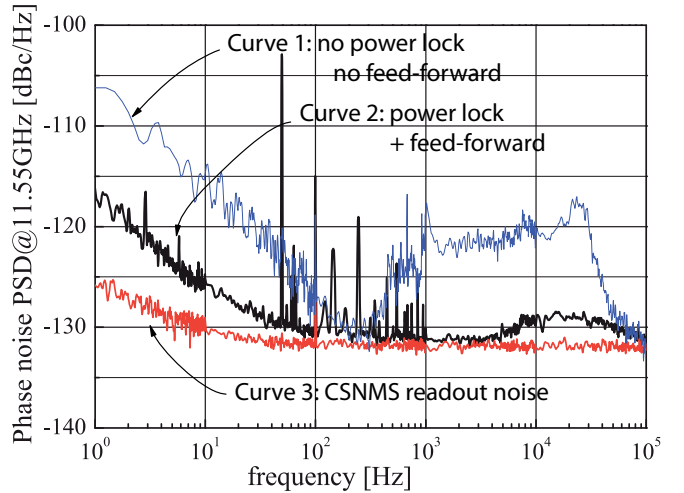}
\caption{Phase noise power spectral density of the difference of the two 11.55\,GHz microwave signals extracted from FOFCs. Assuming identical uncorrelated behavior of the two systems, the phase noise for a single FOFC is 3\,dB lower. Curve 1: No noise reduction technique was applied ; Curve 2: with VCPS in-loop correction and stabilization of the optical power incident on the microwave generating photodiode (both AOM and pump power feed-back give the same result) ; Curve 3: readout system floor.}
\label{fig:Lf}
\end{figure}

Since no carrier is present on the dark port (difference output of the hybrid junction), the associated amplifier and the mixer operate in small-signal regime where flicker noise is greatly reduced \cite{Ivanov_IEEEUFFC_98, Rubiola_RSI_99}. Optimal phase and amplitude tuning produces at least 70 dB of carrier suppression in the dark port leading to a power lower than -100\,dBm before amplification. The bright port (sum output of the hybrid junction) is used to synchronously demodulate the dark port at the carrier frequency, thanks to the mixer and low-pass filter. The phase noise added in this port is therefore common mode and does not impact the final measurement. The readout noise is measured by replacing the input of the dark port by a 50\,$\Omega$ termination (see figure \ref{fig:Lf}, curve 3).

On short timescales, we used a FFT analyzer to measure phase noise power spectral densities of the microwave generation process. On longer timescales, we characterized it by the Allan standard deviation (fractional frequency stability), measured by recording the phase at the output of the CSNMS with a digital voltmeter over extended period of time. We strongly emphasize that under no circumstances did we use any kind of active stabilization (automatic or human-based) of the amplitude or phase matching of the CSNMS. Our carrier suppression system proved perfectly stable over more than several hours of continuous measurement.

%---------------------------------------------------------------------
% VCPS
%---------------------------------------------------------------------

We initially realize a phase comparison between microwave signals directly generated by two FOFCs (see figure \ref{fig:Lf}, curve 1, basically identical whether we implement the CSNMS or a classical detection like in \cite{Millo_OL_09}). A first striking figure of this measurement is a relatively large noise bump between 300 Hz and 80 kHz. This figure is due to the limited servo-bandwidth ($\sim 20$\,kHz) available on one of the FOFCs. We have addressed successfully this issue by implementing, on this FOFC, a feed-forward correction on the microwave using the residual in-loop error signal of the phase-lock-loop. A voltage controlled phase shifter (VCPS) is inserted after the fast photodetector PD 2 associated with the low correction bandwidth FOFC. The control input of the VCPS is driven by a signal derived from the in-loop residual error through a simple voltage divider resistive network. The division ratio is chosen so as to realize a phase ratio between the phase-lock loop residual error signal and the microwave VCPS correction equal to (194\,THz/64)\,/\,11.55\,GHz, with 194\,THz and 11.55\,GHz being respectively the optical and microwave frequencies, the factor 64 being due to the frequency divider in the phase-lock loop. Choosing the right correction sign, this corrects from residual noise due to insufficient gain in the main phase-lock loop. In a sense, this technique is an analog equivalent to the ``transfer oscillator'' scheme, developed by the PTB group \cite{Telle_APB_01}, which uses direct digital synthesizers (DDS) to digitally realize the correct division factor. This technique proved useful to considerably reduce the phase noise spectral density above 300\,Hz as can be seen in figure \ref{fig:Lf} curve 2.

%---------------------------------------------------------------------
% Power stabilization
%---------------------------------------------------------------------

A second striking feature of the noise spectral density in figure \ref{fig:Lf} (curve 1) is the flicker behavior between 1\,Hz and $\sim 200$\,Hz. Studies on Titanium Sapphire based optical frequency combs at NIST \cite{McFerran_EL_05, Bartels_OL_05, Ivanov_UFFC_07} demonstrated that this behavior near carrier was due to AM-PM conversion in the photodetection process. We have implemented two different power stabilization techniques. In both cases, the DC output of the bias-tee following the fast photodiode is compared with a stable voltage reference and sent to an integral analog corrector. In the first technique, the correction signal is applied to the RF-power driving an acousto-optic modulator (AOM, not represented on fig. \ref{fig:SetUp_2FOFC_2PhD}) added in front of the photodiode. In the second technique, the correction is applied to the pump power controller of the FOFC laser oscillator (which is possible because the main phase-lock correction is taken care of by the PZT below a few kHz). With both techniques, the power servo bandwidth is larger than 1\,kHz and is measured (in a separate out-of-loop measurement) to provide more than 20\,dB noise rejection at 1\,Hz Fourier frequency. 

\begin{figure}[ht]
\includegraphics[width=\columnwidth]{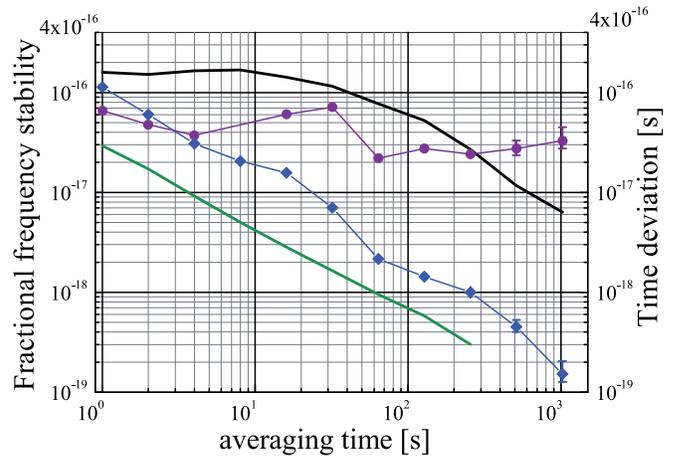}
\caption{Fractional frequency stability (FFS, measured by Allan standard deviation) and time deviation (TDEV) for a single optics-to-microwave FOFC system. Top continuous line: FFS with power stabilization on AOM ; diamonds and circles: respectively FFS and TDEV with power stabilization on pump current control ; bottom continuous line: FFS floor of the measurement system.}
\label{fig:Allan}
\end{figure}

Both power stabilization schemes improved substantially the flicker noise behavior (as seen on figure \ref{fig:Lf} (curve 2)), allowing us to reach an unprecedented phase noise of -117\,dBc/Hz at 1\,Hz from the 11.55\,GHz microwave carrier (-120\,dBc/Hz for a single system). On longer timescales, the measured fractional frequency stabilities were, however, different for the two power stabilization techniques. The AOM-based scheme flattens out up to 10\,s (see figure \ref{fig:Allan}). On the contrary, the pump-power-based technique averages down quickly, almost following the $\tau^{-1}$ slope expected for $1/f$-noise limited phase coherent systems. We believe that the power stabilization of the optical pulses in the oscillator itself has a collateral stabilization effect on their spectral phase (which can be coupled to pulse energy by, for example, Kerr effect or temperature changes), as well as decreasing amplitude to phase conversion in the carrier-envelop offset measurement unit. A thorough and complex investigation would however be necessary to confirm these hypothesis.

In the pump-power stabilization case, the measured fractional frequency stability reaches $1.5 \times 10^{-19}$ at only 1000\,s (for a single system) almost two orders of magnitude better than any previously published data. The level of synchronization between the optics (i.e. the ultra-stable CW common reference laser) and one of the microwave signal is characterized by the time deviation (TDEV) associated with these phase measurements. We find the TDEV to be consistently below 100 attoseconds between 1\,s and 1000\,s, and even reaching 22\,as at 1 minute integration time (see figure \ref{fig:Allan}). This demonstrates the potentiality of our system to become a building block of ultra-high stability time transfer network.

%---------------------------------------------------------------------
% Conclusion
%---------------------------------------------------------------------

Further improvement close to the carrier may involve specific developments in fast photodetector technology with lower $1/f$-noise behavior. Alternatively, opto-electronics detection scheme may be an interesting technique \cite{Kim_Nature08}. At high Fourier frequencies, improving the white noise background would require a better signal to noise ratio which may be accessible by increasing the saturation power of the photodetector and/or multiplying the repetition rate. Although it seems hard with the present FOFC technology to go much beyond a few hundreds MHz repetition rate (except with harmonically mode-locked lasers), pulse multiplication by resonant Fabry-Perot cavity \cite{Kirchner_OL_09, Steinmetz_APB_09} is a very promising technology which may reasonably be implemented for our FOFCs.

%---------------------------------------------------------------------
% Bibliography
%---------------------------------------------------------------------

%---------------------------------------------------------------------
% Figures
%---------------------------------------------------------------------

% Only for separate figures pages...

\end{document}